\documentstyle[11pt,paspconf,epsf]{article}

\begin{document}

\title{Galaxy Clusters and Large Scale Structure at High Redshifts}

\author{Paul J. Francis\altaffilmark{1}}
\affil{Mt. Stromlo and Siding Spring Observatory, Private Bag, Weston Creek 
Post Office, Weston, ACT 2611, Australia \\ E-mail: pfrancis@mso.anu.edu.au}

\altaffiltext{1}{Joint appointment with the Department of Physics and
Theoretical Physics, Australian National University}

\author{Bruce E. Woodgate}
\affil{NASA Goddard Space-Flight Center, Code 681, Greenbelt, MD 20771, 
USA, E-mail: woodgate@achamp.gsfc.nasa.gov}

\author{Anthony C. Danks}
\affil{Hughes STX, Goddard Space-Flight Center, Code 683.0, Greenbelt, 
MD 20771, USA, E-mail: danks@iue.gsfc.nasa.gov}

\begin{abstract}
We present a detailed study of a rich galaxy cluster at z=2.38. We
demonstrate that this cluster contains large overdensities of
damped Ly$\alpha$ absorption lines, of Ly$\alpha$ emitting galaxies
and of extremely red objects. The overdensity of extremely red
objects in this field demonstrates that many are high $z$ galaxies.

The huge overdensities we measure for these three classes of object are
much larger than the {\em mass} overdensities of typical clusters at this
redshift, as predicted by CDM and related models. We suggest therefore
that the distribution of damped Ly$\alpha$ absorption line systems, of 
Ly$\alpha$ emitting galaxies
and of extremely red objects are all very strongly
biassed, and that somehow a small overdensity of mass has increased
the fraction of baryons in collapsed objects, in the volume occupied by 
the cluster, to close 
to unity (a factor of $\sim  10$ increase).

We speculate that some unknown physical process, acting on the volume
occupied by our cluster, caused the normally diffuse ionised
inter-galactic medium to coalesce into small ($< 10^8 M_{\sun}$)
blobs of neutral hydrogen, which produce the Ly$\alpha$  absorption-lines. Star
formation occurred within these blobs at $z>5$, enriching them with 
metals and producing stars, which after several mergers and $\sim 0.5$ Gyr
of passive evolution form the extremely red objects. The Ly$\alpha$
emitting galaxies are probably AGN, triggered perhaps by mergers of the
small blobs.

\end{abstract}

\keywords{Galaxies: clusters: formation: individual (2139-4434) -- Quasars:
absorption-lines}

\section{Introduction}

It is now abundantly clear that many classes of object in the young
universe are very strongly clustered (eg. Cohen, Cristiani, Dickinson,
Postman and Steidel in these proceedings). This contrasts strongly with 
theoretical models, which predict that matter should be very weakly
clustered at high redshifts (eg. Brainerd \& Villumsen 1994). As
described by Kauffmann and Matarrese in these proceedings, the
apparent contradiction can be resolved by strong biassing in the young
universe: at high redshift, galaxies may only have formed in the highest
density regions (which are strongly clustered), despite the relatively
modest matter concentrations in these regions.

It is less clear what physical mechanism could be responsible for the biassing.
Why should modest ($\sim 10$\%) enhancements in the mass density trigger
enormous ($ > 100$\%) enhancements in the galaxy density? Most simulations
use ad-hoc approximations, such as threshold densities below which no
star formation can occur: such arbitrary assumptions can crudely reproduce
the clustering seen, but have no strong physical motivation. 

\begin{figure}

\plotone{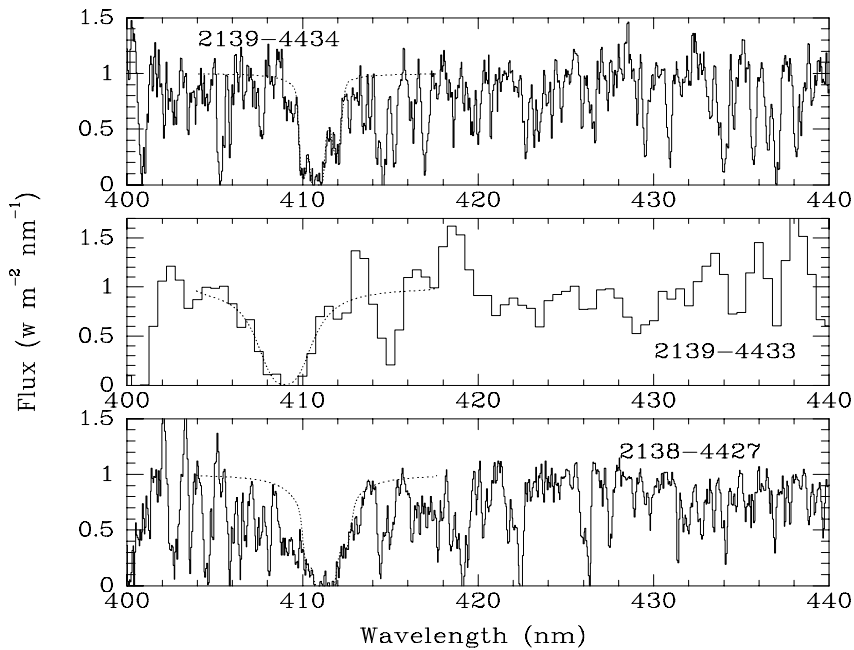}

\caption{Spectra of three $z \sim 3.2$ QSOs lying behind the cluster,
showing the Ly$\alpha$ absorption-line systems at $z\sim 2.38$. The
dotted lines show Voigt profile fits to the spectra: note that multiple
components are required. The spectrum of 2139-4433 is of much lower
spectral resolution than the others.}

\end{figure}

In an attempt to come to a more physical
understanding of the biassing process, we have
focussed on perhaps the best studied cluster in the early universe: the
2139$-$4434 cluster at z=2.38 first identified by Francis \& Hewett 
(1993). In this paper, we present preliminary results of an inventory of 
the constituents of the cluster. A full account of this work is
in preparation (Francis, Woodgate \& Danks 1998, in preparation)

\section{QSO Absorption-lines}

The cluster was first identified in absorption against two background
QSOs. We now have spectra of three QSOs lying behind this cluster, and
all three show Ly$\alpha$ absorption-line systems at $z=2.378 \pm 0.015$
with neutral hydrogen column densities $N_H \sim 10^{20} {\rm cm}^{-2}$
(Fig~1). The three QSOs lie roughly along a line in the sky, with
the furthest two separated by 8$^{\prime}$, which at $z=2.38$ corresponds
to a comoving transverse separation between the lines of sight of
$\sim 10$ co-moving Mpc.

The probability of finding three absorption systems with these column 
densities within such a small volume by chance are tiny: the average
number of Ly-limit absorbers per unit redshift at $z \sim 2.4$ is $\sim 2$, 
but we have three within
a pathlength $\Delta z < 0.05$, implying an overdensity factor $d \sim 30$,
with a 95\% confidence lower-limit of $d > 7.5$. We therefore conclude
that the region of space, roughly 10 Mpc in size, lying in front of 
these QSOs at $z \sim 2.38$,
is very significantly overdense in Ly-limit absorption-line systems.

\begin{figure}

\plotone{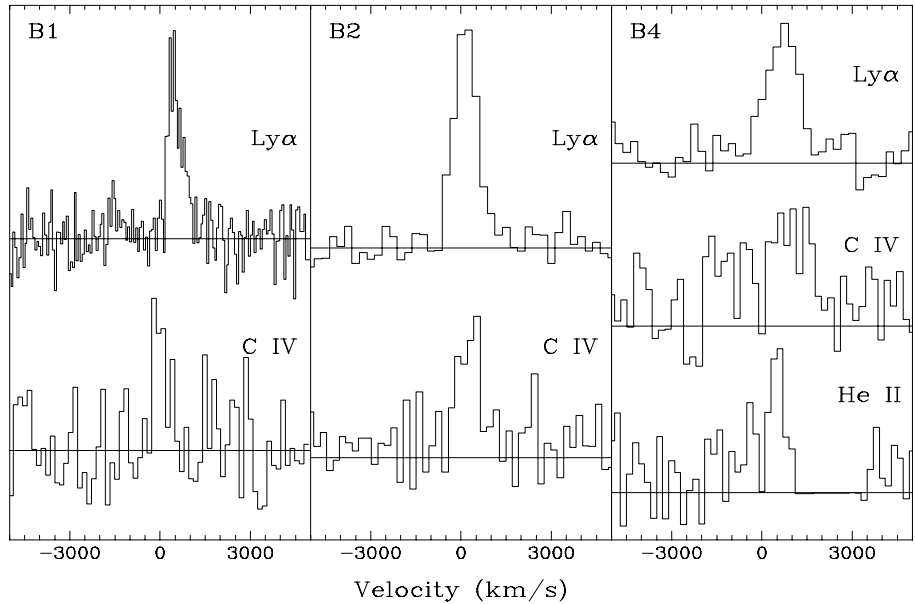}

\caption{Spectra of three Ly$\alpha$-emitting galaxies at $z=2.38$ lying 
within the absorption-line cluster.}

\end{figure}

High resolution spectroscopy of the absorbers shows that they are
made up of multiple components, each with velocity dispersions of 
$> 100 {\rm km\ s}^{-1}$. Several of the components show metal-line 
absorption: the data is too poor to determine metalicities, but the
line ratios suggest that the Hydrogen is predominantly neutral, as
would be expected from the column densities.

\section{Ly$\alpha$ Emitting Galaxies}

We carried out a narrow-band search for Ly$\alpha$ 
emission associated with the cluster of QSO absorption-lines 
(Francis et al. 1996, Francis, Woodgate \& Danks 1997). Three 
powerful sources
were detected ($F_{\rm Ly\alpha}
\sim  5 \times 10^{-16} {\rm erg\ cm}^{-2}{\rm s}^{-1}$). All three
have spectral confirmation (Fig~2); on the basis of their line ratios, 
velocity widths and equivalent widths, all three are probably narrow-line
AGNs.

\begin{figure}

\plotone{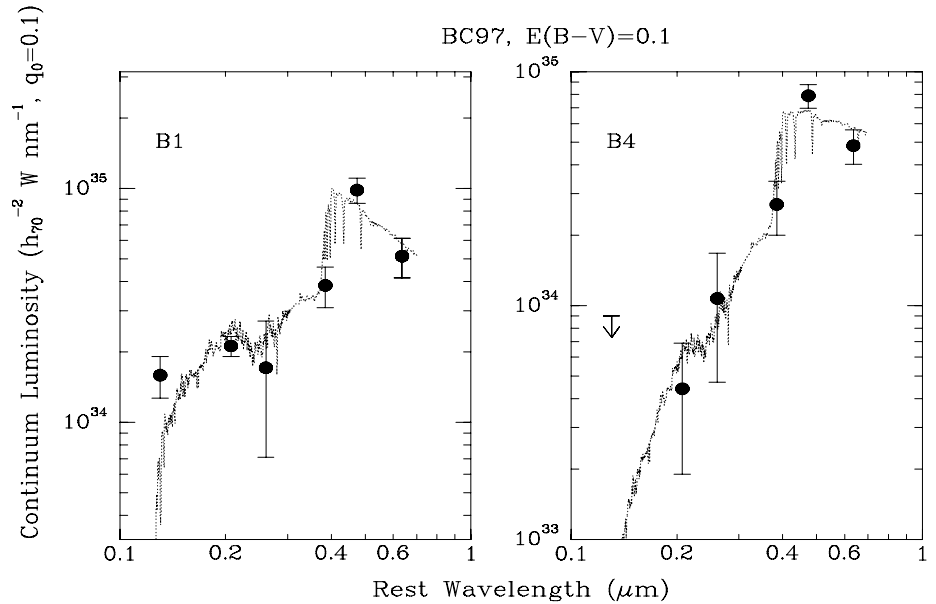}

\caption{Continuum spectral energy distributions of two EROs from our
field with confirmed spectroscopic redshifts, showing the extremely
red colours. For comparison, the spectral energy distributions of
an instantaneous stellar burst population observed $5\times 10^8$ years
after formation and obscured by a small amount ($E(B-V) \sim 0.1$) of
dust are shown  (Bruzual \& Charlot 1997, in preparation).}

\end{figure}

We detect three Ly$\alpha$ emitting galaxies in a co-moving volume of 
460 Mpc$^3$, while
Mart\'{\i}nez-Gonz\'alez et al. (1995) surveyed a
co-moving volume of 1400 Mpc$^3$ at $z=3.4$ for Ly$\alpha$ emitting sources 
to a comparable flux limit but detected nothing. An overdensity of Ly$\alpha$
emitting galaxies in our field of $d > 3$ (on a scales of $\sim 10$
co-moving Mpc) is required to raise the joint probability of seeing three
galaxies in our field, and seeing none in Mart\'{\i}nez-Gonz\'alez et al.'s 
data, to 5\%. Thus the volume of space containing the overdensity of QSO
absorption systems is also significantly overdense in Ly$\alpha$ emitting
galaxies.

\section{Extremely Red Objects}

Many near-IR imaging surveys are detecting extremely red objects (say
$R-K >5$) with $K \sim 20$ (eg. Meisenheimer and Waddington, these
proceedings). At least some of the extremely red objects (EROs) are
high redshift galaxies (eg. Graham \& Dey 1996).

We imaged the central 4 square arcmin of the cluster and detected
6 EROs down to $K_n \sim 20.5$. Meisenheimer et al.
surveyed 120 square arcmin in $R$ and $K$ to a similar flux limit
and discovered 20 EROs. This implies that our field has an overdensity
factor of EROs $d \sim 9$, with a 95\% lower limit of $d > 2.5$. This
overdensity in a field chosen to contain a high redshift cluster
{\em confirms that most EROs are red, high redshift galaxies}. Indeed, 
three of the EROs in our field have detectable Ly$\alpha$ and/or
H$\alpha$ emission, which confirms that they lie at $z \sim 2.38$.

Why are these high redshift galaxies so red? We have new 
multi-colour photometry of two of the EROs with confirmed redshifts,
suggesting that the redness is caused by a red-shifted 4000\AA\ break
lying between the $J$ and $H$ bands
in the spectral energy distributions (Fig~3); this implies that the
EROs are massive ($ M_{\rm baryonic} \sim 5 \times 10^{11} M_{\sun}$),
and that they completed the bulk of their star formation at least $5 \times
10^8$ years before they are observed (ie. at $z>5$). Alternative 
explanations, such as dusty starbursts, fail to reproduce the colours
observed, in particular the extremely red $J-H$ colours.

\section{Discussion}

The $z=2.38$ cluster is $> 10$ co-moving Mpc long, $> 1$ Mpc wide, and 
if we interpret the redshift dispersion of cluster members as caused by 
the Hubble flow rather than peculiar motions, $\sim 10$ Mpc deep. We thus
have to explain how a region of the universe roughly 10 Mpc in size
can have overdensities of neutral hydrogen clouds, AGN and red galaxies
of $> 7$, $> 3$ and $> 2.5$ respectively. We cannot use the Ly-break
search technique at this redshift, but for all we know this region could
have a substantial overdensity of Ly-break galaxies as well.

It is a generic prediction of CDM and related models that {\em mass} will
be far less clustered at $z \sim 2.4$ than at present; mass fluctuations on
scales of 10 co-moving Mpc are expected to be $\sim \times 10$ smaller
than at present: ie. rms mass fluctuations should be around 10\%. This
is clearly far smaller than the observed overdensities. If the standard
model is correct, the overdensities must therefore be due to biassing: the
fraction of matter which is in detectable forms must be greatly enhanced
in this volume.

\subsection{Baryon Inventory}

Are there sufficient baryons in the cluster volume to form all the observed
objects? If we assume that the cluster is 10 co-moving Mpc deep, then
the baryonic mass available per square comoving Mpc ($\sim$ 1 arcmin square)
is $7 \times 10^{10} (\Omega_{\rm baryon}/0.05) 
(H_0/70 {\rm km\ s}^{-1}{\rm Mpc}^{-1})^2 M_{\sun}{\rm Mpc}^{-2}$
(for $q_0 = 0.1$). If we 
assume that the neutral hydrogen clouds have a covering factor of $\sim 1$, 
their baryonic mass surface density is $5 \times 10^{10} 
M_{\sun}{\rm Mpc}^{-2}$. The baryonic mass surface density in red galaxies
can be calculated from the spectral synthesis models. Assuming a Salpeter
IMF and placing an upper limit on the dust extinction derived from the 
observed Ly$\alpha$/H$\alpha$ ratios gives a surface density of
$> 10 \times 10^{10} M_{\sun}{\rm Mpc}^{-2}$ (for $H_0 = 70 {\rm km\ 
s}^{-1}{\rm Mpc}^{-1}$ and $q_0 = 0.1$).

{\em Thus the baryonic mass in both the observed absorption-lines and in the
red galaxies is at least comparable to the total baryonic mass
available}. Tinkering with the gas covering factor or the galaxy stellar
initial mass function can bring the measured baryonic masses down
somewhat, but it seems probable that most 
of the baryons in the volume have been incorporated into the sources
observed. Note that we are insensitive to Ly-break galaxies
and to red galaxies fainter than $\sim 3 \times L_*$: these too may add to
the baryonic mass present.

The baryonic mass density we determine is an order of magnitude
larger than the typical baryonic mass density found in 
damped Ly$\alpha$ systems and Ly-limits absorbers at this redshift
(Lanzetta et al. 1991, Steidel 1990). The reservoir from which the
baryons came can only therefore be the Ly$\alpha$ forest and ionised
phase of the inter-galactic medium (IGM), which can contain most of 
the baryonic matter in the universe at this redshift (eg. Miralda-Escude 
et al. 1996). The puzzle therefore is what caused a large fraction of 
the IGM to turn into neutral hydrogen clouds, red galaxies and AGN.

\subsection{Why is the Hydrogen Neutral?}

The baryons in the IGM and Ly$\alpha$ forest are very highly ionised:
why then should a large fraction have been converted into neutral
hydrogen clouds in this region of space, forming the absorption-line
cluster?

One possibility is that there are simply no UV sources
close enough to have photoionised this volume. The mean free path of ionising 
photons at this redshift is, however, $\sim 500$ Mpc, and one would
expect to find hundreds of QSOs within such a volume: indeed, we know of
at least 8 luminous QSOs within 500 Mpc of the absorption-line cluster.
Simulations confirm that fluctuations in the ionising flux due to the
random distribution of ionising sources are likely to be small at these
redshifts, due to the averaging effect of the large mean free paths
(eg. Croft 1997).

A second possibility is that a small overdensity of Ly-limit systems
shields a volume of space from the ionising flux, causing it to 
become neutral. The overdensity needed to make shielding significant on
10 Mpc scales is however $d \sim 50$; comparable to the measured
overdensity of the cluster. Thus while the cluster may be
dense enough to screen the IGM within it from ionising photons, this
screening cannot explain why it is so dense.

As variations in the ionising flux cannot account for the large
neutral gas mass of the cluster, it must be caused by the physical state
of the gas. An increase in gas density by a factor of $> 10^4$ is
necessary to decrease the ionisation potential enough to make the gas 
predominantly neutral. If the IGM is compressed into small blobs filling
$10^{-4}$ of the volume, the blobs must be small ($< 10$ kpc): larger blobs
have too small a ratio of surface area to mass to bring the absorption-line
covering factor of the cluster up to $\sim 1$, as required by the 
absorption-line statistics.

To explain the absorption-line properties, we thus require that most of the
IGM has somehow formed small ($< 10^8 M_{\sun}$) dense clouds filling less
that $10^{-4}$ of the cluster volume.

\subsection{What are the Red Galaxies?}

The extremely red galaxies in our cluster have the colours and magnitudes
of old ($> 5 \times 10^8$ years), massive ($> 10^{12} M_{\sun}$, if
dark matter is included) elliptical galaxies. It is not possible to
form objects this massive at sufficiently high redshifts in most
cosmological models (eg. Kashlinsky \& Jiminez 1997). Kauffmann (1996)
proposed a solution to this problem in the context of elliptical
galaxy formation: the stellar population could form in many small
blobs at high redshift, which merge at a later stage to form a
single massive red galaxy. A drawback of this model is that it 
requires that galaxies collide without triggering significant star formation.

We can thus possibly explain the red galaxies by assuming that in
the co-moving volume now occupied by the cluster, many low-mass
galaxies formed at $z>5$, perhaps in the same small neutral gas blobs
that we are seeing in absorption. These blobs are continuously merging,
and in a few places, sufficient blobs have accumulated to form red
galaxies that make it over our detection threshold. The frequent mergers
of the red blobs could perhaps explain the overdensity of AGN in
the cluster.

\subsection{An Extremely Tentative Model}

Here then is our current speculation about the origin of this cluster.
At $z>5$, some unknown factor causes most of the IGM in this 
$\sim 10$ Mpc scale volume to collapse into many small 
($\sim 10^6 M_{\sun}$?) dense clumps. Stars form in these clumps,
which then subsequently merge, triggering AGN activity (but not
further star formation) and slowly assembling large red galaxies.
By $z=2.38$, a few massive red galaxies have been assembled, but
most of the gas blobs either contain no galaxy, or a galaxy below
our detection threshold, and can only be detected in absorption.
By $z = 0$, the merger process will have continued, and produced
a full complement of cluster ellipticals, while the remaining 
neutral gas is now so hot that it emits X-rays, though still metal
enriched by the high redshift star formation. This structure
would thus be the ancestor of a rich galaxy cluster today.

The puzzle remains: what caused the IGM in this particular
part of the universe to collapse into small blobs in the first place?
The real question may however be the opposite: why didn't
this collapse take place in the rest of the universe? Clouds of
$\sim 10^6 M_{\sun}$ have masses far below
the Jeans mass even at recombination: the usual problem in
cosmological simulations is not how to form them, but how to prevent 
most of the baryons in the universe from collapsing into such blobs
long before $z \sim 100$
(eg. White \& Rees 1978). This delay is generally thought to be caused 
by stellar feedback: energy generation by the first stars to form in 
these clumps aborts further star formation, perhaps by blowing the
gas out of the shallow potential wells.

Our cluster can therefore be explain by some mechanism that inhibited
the stellar feedback that, in most of the universe, shut down or
delayed proto-galaxy formation at $z>5$. Somehow, a small ($\sim 5\%$
at $z \sim 5$?) mass overdensity must dramatically alter the normal
processes of galaxy formation. We have no idea how!

\section{Conclusions}

\begin{itemize}

\item Most of the baryons in
a high redshift cluster are either in the form of an old stellar population,
or in small dense blobs of neutral gas. Somehow, the small 
enhancements of mass density expected in high redshift clusters 
have greatly increased
the fraction of the ionised IGM that is in the form of compact neutral
clumps.

\item Much of the gas in at least one rich galaxy cluster 
at $z=2.38$ is neutral and distributed in small, metal enriched blobs.

\item Many extremely red objects are high redshift galaxies.

\end{itemize}

\acknowledgments

We wish to thank Bruzual and Charlot for letting us use the results of 
their latest spectral synthesis codes in advance of publication, 
and V\'eron and Hawkins for showing us their discovery spectrum of
QSO 2139-4433.


\begin{references}
\reference Brainerd, T. G., \& Villumsen, J. V. 1994, \apj , 431, 477
\reference Croft, R. A. C. 1997, preprint (astro-ph/9709302)
\reference Francis, P. J., \& Hewett, P. C. 1993, \aj, 105, 1633
\reference Francis, P. J. et al. 1996, \apj , 457, 490
\reference Francis, P. J., Woodgate, B. E., \& Danks, A. C., 1997, \apj ,
482, L25
\reference Graham, J. R., \& Dey, A. 1996, \apj , 471, 720
\reference Kauffmann, G. 1996, \mnras , 281, 487
\reference Kashlinsky, A., \& Jiminez, R. 1997, \apj , 474, L81
\reference Lanzetta, K. M., McMahon, R. G., Wolfe, A. M., Turnshek, D. A.,
Hazard, C., \& Lu, L., 1991, \apjs , 77, 1
\reference Mart\'{\i}nez-Gonz\'alez, E., Gonz\'alez-Serrano, J. I.,
Cay\'on, L., Sanz, J. L., \& Mart\'{\i}n-Mirones, J. M. 1995, \aap , 303, 379
\reference Miralda-Escude, J., Cen, R., Ostriker, J. P., \& Rauch, M., 1996,
\apj , 471, 582
\reference Steidel, C. C. 1990, \apjs , 74, 37
\reference White, S. D. M., \& Rees, M. J., 1978, \mnras , 183, 341

\end{references}
\end{document}